\documentclass[aps,pre,nofootinbib,showpacs,twocolumn]{revtex4-1}
\usepackage{amssymb,latexsym,mathrsfs}
\usepackage{amsfonts}
\usepackage{mathrsfs}
\usepackage{amsmath}
\usepackage{graphicx}
\usepackage{color}

\newcommand{\beq}{\begin{equation}}
\newcommand{\eeq}{\end{equation}}
\newcommand{\beqn}{\begin{eqnarray}}
\newcommand{\eeqn}{\end{eqnarray}}
\newcommand{\bearr}{\begin{array}}
\newcommand{\enarr}{\end{array}}

\def\bea{\begin{eqnarray}}
\def\eea{\end{eqnarray}}
\def\ba{\begin{array}}
\def\ea{\end{array}}

\begin{document}
\title{Role of dimensionality in complex networks: Connection with nonextensive statistics}
\author{S.G.A. Brito$^1$}
\email[E-mail address: ]{samuraigab@dfte.ufrn.br}
\author{L.R. da Silva$^{1,2}$}
\email[E-mail address: ]{luciano@dfte.ufrn.br}
\author{Constantino Tsallis$^{2,3}$}
\email[E-mail address: ]{tsallis@cbpf.br}
\affiliation{$^1$Departamento de F\'{\i}sica Te\'orica e Experimental, Universidade Federal do Rio Grande do Norte, Natal, RN, 59078-900, Brazil}
\affiliation{$^2$National Institute of Science and Technology of Complex Systems, Brazil}
\affiliation{$^3$Centro Brasileiro de Pesquisas Fisicas, Rua Xavier Sigaud 150, 22290-180 Rio de Janeiro-RJ,  Brazil \\
and 
Santa Fe Institute, 1399 Hyde Park Road, New Mexico 87501, USA }
%\date{\today}

\begin{abstract}
Deep connections are known to exist between scale-free networks and non-Gibbsian statistics. For example, typical degree distributions at the thermodynamical limit are of the form $P(k) \propto e_q^{-k/\kappa}$, where the $q$-exponential form $e_q^z \equiv [1+(1-q)z]^{\frac{1}{1-q}}$  optimizes the nonadditive entropy $S_q$ (which, for $q\to 1$, recovers the Boltzmann-Gibbs entropy). We introduce and study here $d$-dimensional geographically-located networks which grow with preferential attachment involving Euclidean distances through $r_{ij}^{-\alpha_A} \; (\alpha_A \ge 0)$. Revealing the connection with $q$-statistics, we numerically verify (for $d$ =1, 2, 3 and 4) that the $q$-exponential degree distributions exhibit, for both $q$ and $\kappa$, universal dependences on the ratio $\alpha_A/d$. Moreover, the $q=1$ limit is rapidly achieved by increasing $\alpha_A/d$ to infinity. 
\end{abstract}

\pacs{89.75.Hc, 05.70.-a, 05.45.Pq, 89.75.Da}

\maketitle

Networks emerge spontaneously in many natural, artificial and social systems. Their study is potentially important for physics, biology, economics, social sciences, among other areas.  For example, many empirical studies have identified peculiar properties in very different networks such as the Internet and online social networks (e.g., Facebook), citations networks, neurons networks~\cite{strogatz2001,newman2003,fontoura2011}, to quote but a few.  An ubiquitous class of such networks is constituted by the scale-free ones (more precisely, asymptotically scale-free).  As we shall soon verify, these networks can be seen as a particular application of nonextensive statistical mechanics, based on the nonadditive entropy $S_q=k\frac{1-\sum_ip_i^q}{q-1} \; (q \in {\cal R}; \, S_1=S_{BG \equiv }-k\sum_i p_i \ln p_i$, where $BG$ stands for {\it Boltzmann-Gibbs}) \cite{Tsallis1988,GellMannTsallis2004}. This current generalization of the BG entropy and corresponding statistical mechanics has been widely successful in clarifying the foundations of thermal statistics as well as for applications in complex systems in high-energy collisions at LHC/CERN (CMS, ALICE, ATLAS and LHCb detectors) and at RHIC/Brookhaven (PHENIX detector) \cite{cs9}, cold atoms \cite{cs1}, dusty plasmas \cite{cs2},  spin-glasses \cite{cs3}, trapped ions \cite{cs4}, astrophysical plasma \cite{cs5}, biological systems \cite{cs7}, type-II superconductors \cite{cs8}, granular matter \cite{cs15} (see \cite{Bibliography}). 

The deep relationship between scale-free networks and $q$-statistics started being explored in 2005 \cite{SoaresTsallisMarizSilva2005,ThurnerTsallis2005,Thurner2005}, and is presently very active \cite{Andradeetal2005,Lindetal2007,Mendesetal2012,Almeidaetal2013,Luciano}. The basic connection comes (along the lines of the BG canonical ensemble) from the fact that, if we optimize the functional $S_q[P(k)]= k\frac{1-\int dk [P(k)]^q}{q-1}$ with the constraint $\langle k\rangle \equiv \int dk\, k P(k)= constant$ or analogous ($k$ being the degree of a generic site, i.e., the number of links that arrive to a given site; $P(k)$ denotes the degree or connectivity distribution), we straightforwardly obtain $P(k) = P(0)e_q^{-k/\kappa}=P(0)/[1+(q-1)k/\kappa]^{\frac{1}{q-1}}$, which turns out to be the generic degree distribution for virtually all kinds of scale-free networks. The $q$-exponential function is defined as $e_q^z \equiv [1+(1-q)z]^{\frac{1}{1-q}} \; (e_1^z=e^z)$. We verify that, for $q>1$ and $k\to\infty$, $P(k) \sim  1/k^\gamma$ with $\gamma \equiv 1/(q-1)$. The classical result $\gamma=3$ \cite{BarabasiAlbert1999} corresponds to $q=4/3$.

In the present work we address the question of how universal such results might be, and more specifically, how $P(k)$ varies with the dimension $d$ of the system? 

\begin{figure}[!htb]
\centering		
\includegraphics[scale=.27]{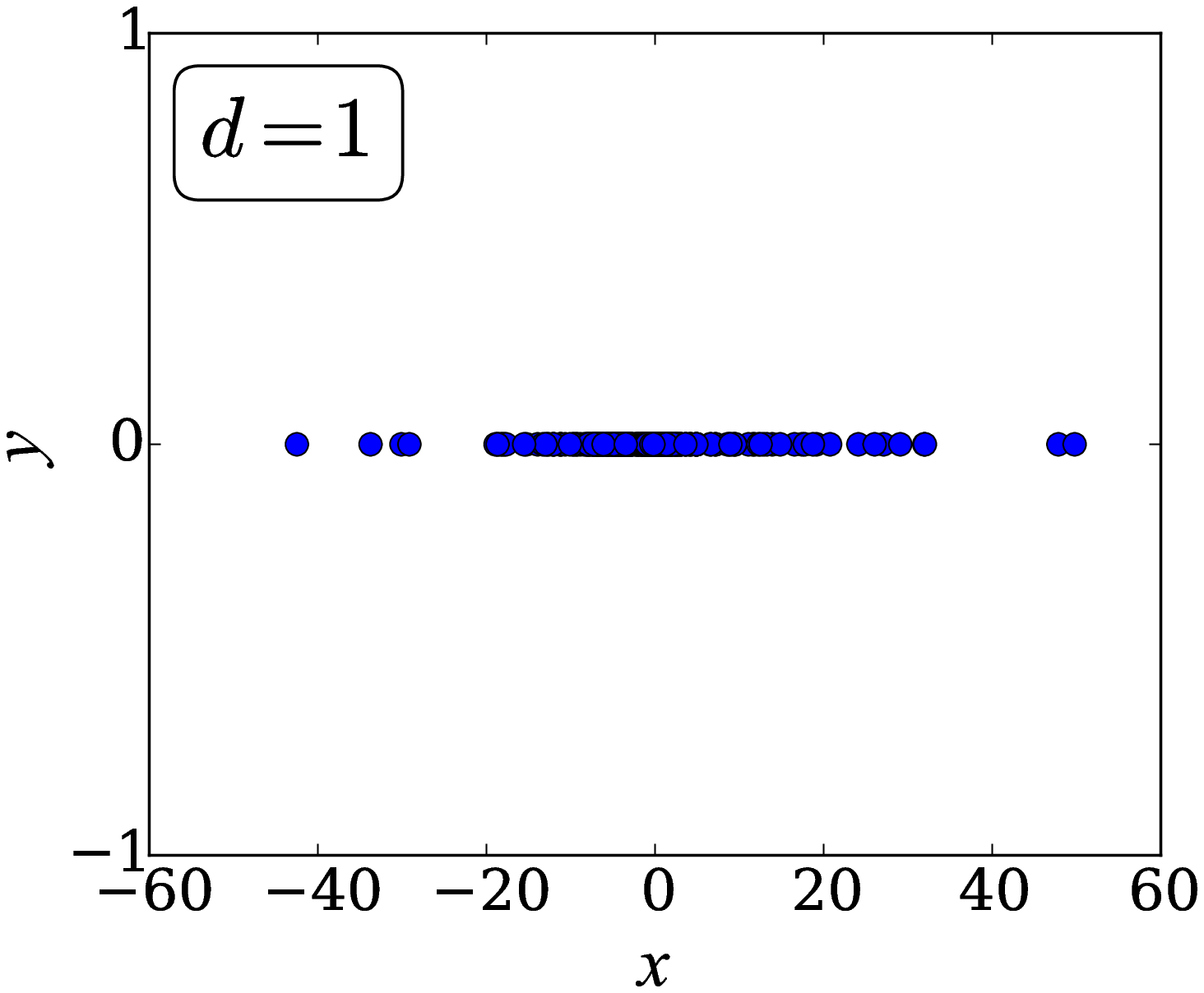} 	
\includegraphics[scale=.27]{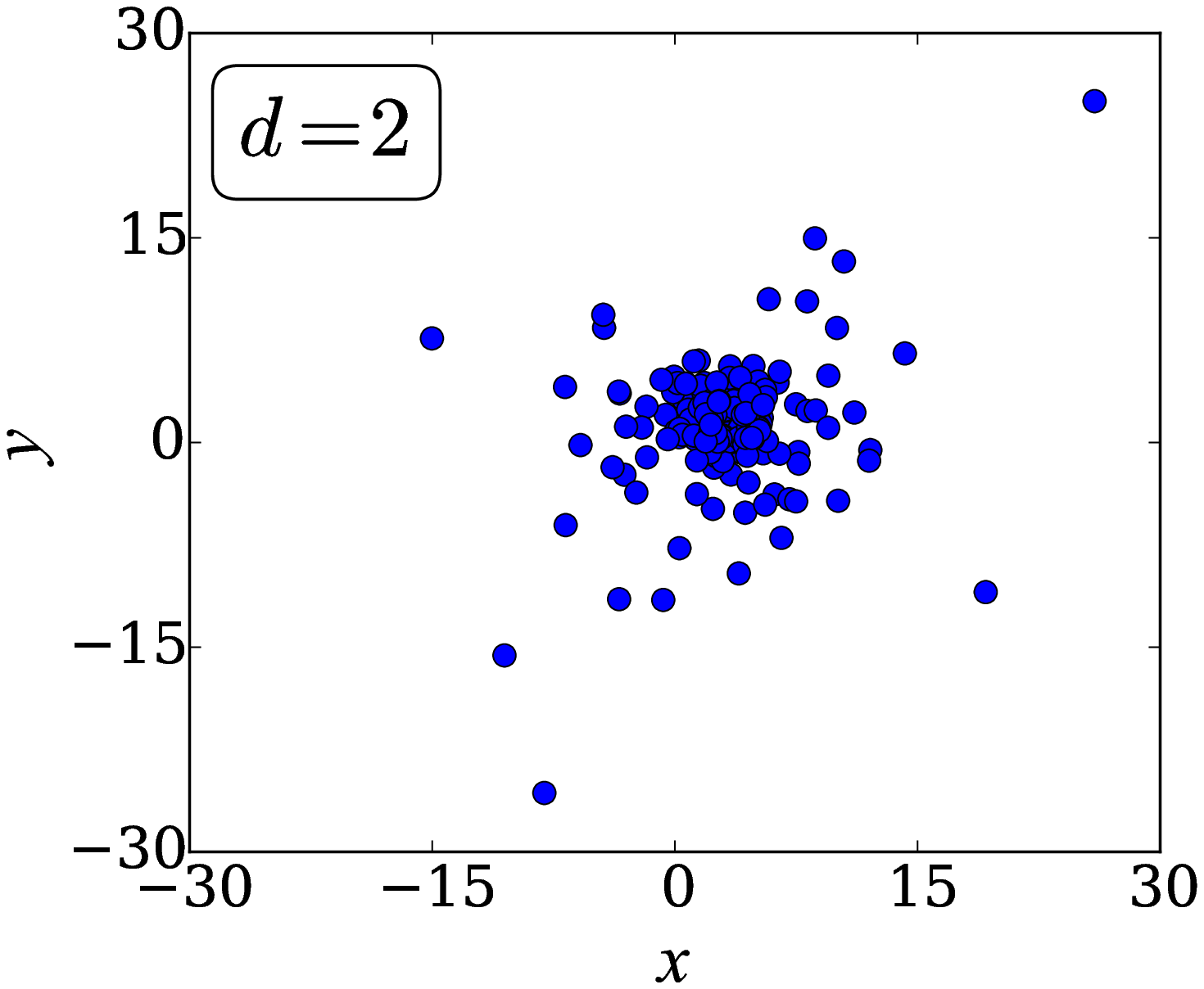} 	
\includegraphics[scale=.38]{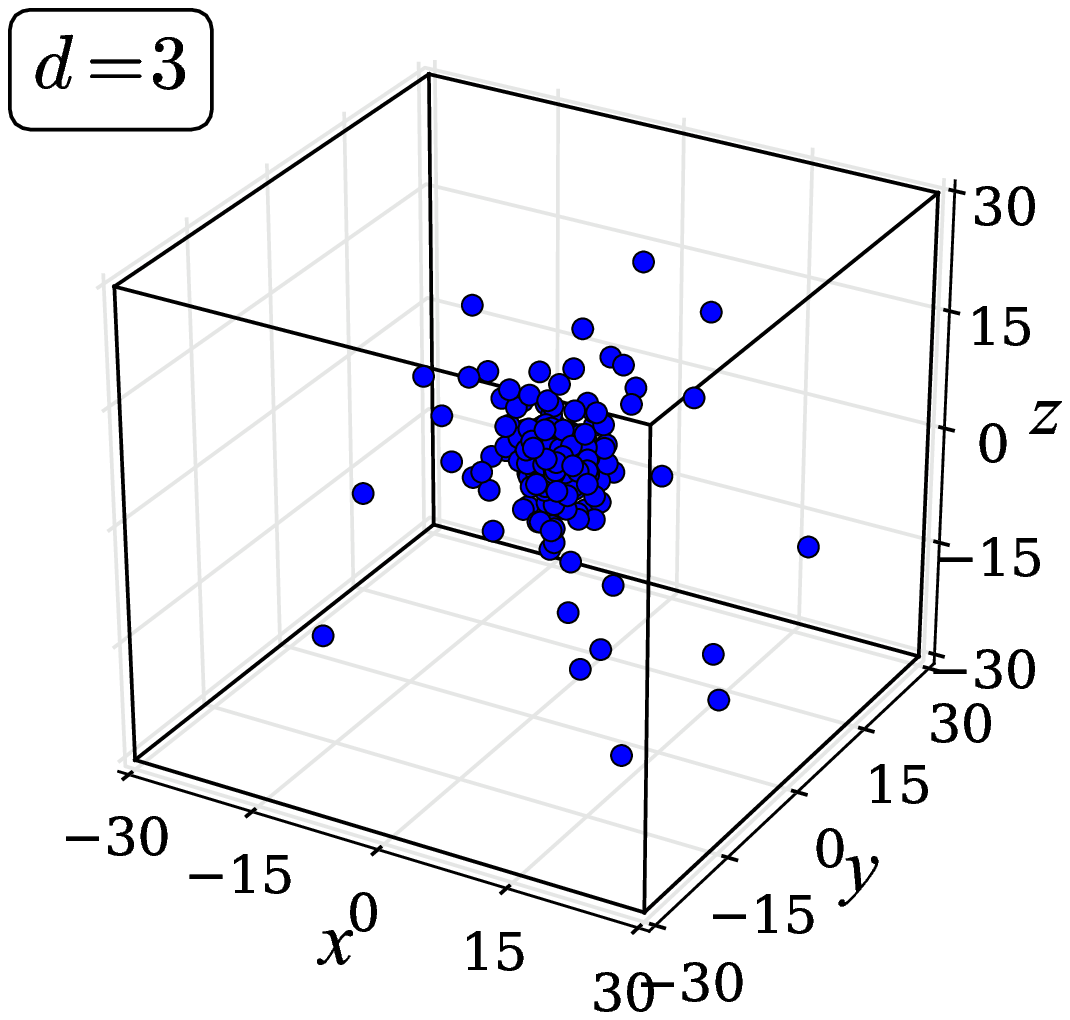} 	 
\caption{Distribution of $N=500$ sites obtained with Eq. (\ref{eq:distance}) for $\alpha_A = 2.0$, $\alpha_G = 0.0$, and $d=1,2,3$. }
\label{sitios}
\end{figure}

Our growing model starts with one site at the origin. We then stochastically locate a second site  (and then a third, a fourth, and so on up to $N$) through the $d$-dimensional isotropic distribution 
\begin{equation} \label{eq:distance}
 p(r) \propto \frac{1}{r^{d+ \alpha_G}}    \;\;\;(\alpha_G >0; \, d=1,2,3,4) \,,
\end{equation}
where $r \ge 1$ is the Euclidean distance from the newly arrived site to the center of mass of the pre-existing system (in one dimension, $r = |x|$; in two dimensions, $r = \sqrt{x^2 + y^2}$; in three dimensions $r = \sqrt{x^2 + y^2 + z^2}$, and so on); $p(r)$ is zero for  $0 \le r<1$; the subindex $G$ stands for {\it growth}. We consider $\alpha_G >0$ so that the distribution $P(r)$ is normalizable; indeed, $\int_1^\infty dr\,r^{d-1} r^{-(d+ \alpha_G)} = \int_1^\infty dr\,1/r^{1+\alpha_G}$, which is finite for $\alpha_G >0$, and diverges otherwise.
 See Fig.\ref{sitios}. 
 
 Every new site which arrives is then attached to one and only one site of the pre-existing cluster. The choice of the site to be linked with is done through the following preferential attachment probability:
 \begin{equation}\label{eq:pdk}
 	\Pi_{ij}= \frac{k_i \,{r_{ij}}^{-\alpha_A}}{{\sum k_i }\,{r_{ij}}^{-\alpha_A}} \in [0,1] \;\;\;(\alpha_A \ge 0)\,,
 \end{equation}
where $k_i$ is the connectivity of the $i$-th pre-existing site (i.e., the number of sites that are already attached to site $i$), and $r_{ij}$ is the Euclidean distance from site $i$ to the  newly arrived site $j$; subindex $A$ stands for {\it attachment}.

For $\alpha_A$ approaching zero and arbitrary $d$, the physical distances gradually loose relevance and, at the limit $\alpha_A = 0$, all distances becomes irrelevant in what concerns the connectivity distribution, and we therefore recover the Barab\'asi-Albert (BA) model~\cite{BarabasiAlbert1999}, which has topology but no metrics.

\begin{figure}[!htb]
\centering
\includegraphics[scale=.28]{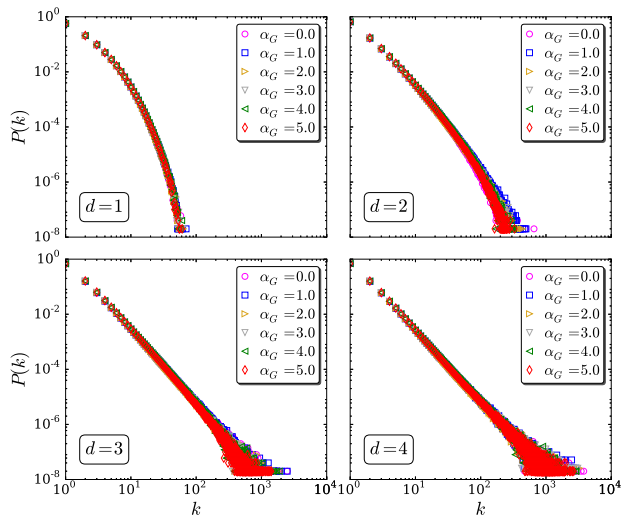} 		 
\caption{Connectivity distribution for $d=1,2,3,4$, $\alpha_A = 2.0$ and typical values for $\alpha_G$. The simulations have been run for $10^3$ samples of $N = 10^5$ sites each. We verify that $P(k)$ independs from $\alpha_G$ ($\forall d$).
}
\label{alfa_g}
\end{figure} 

\begin{figure}[htb]
\centering
\includegraphics[scale=.25]{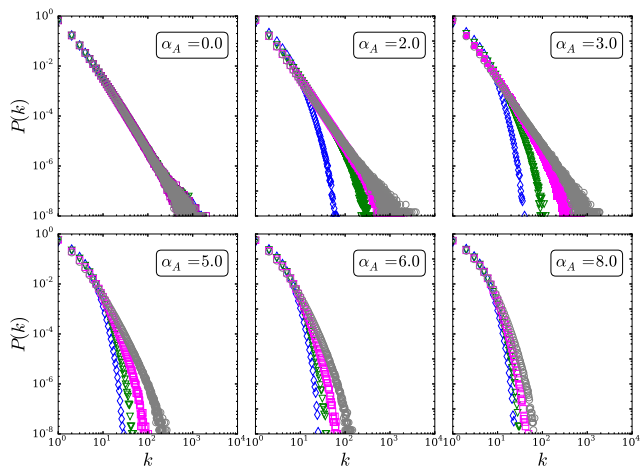}
\caption{Degree distribution for $d=1 \, (blue \,diamonds)$, $2 \,(green \,triangles),3 \,(magenta \,squares),4 \,(grey \,circles)$, and typical values of $\alpha_A$, with $\alpha_G = 2.0$.  The simulations have been run for $10^3$ samples of $N = 10^5$ sites each.}
\label{compare}
\end{figure}

\begin{figure}[!htt]
\centering
\includegraphics[scale=.26]{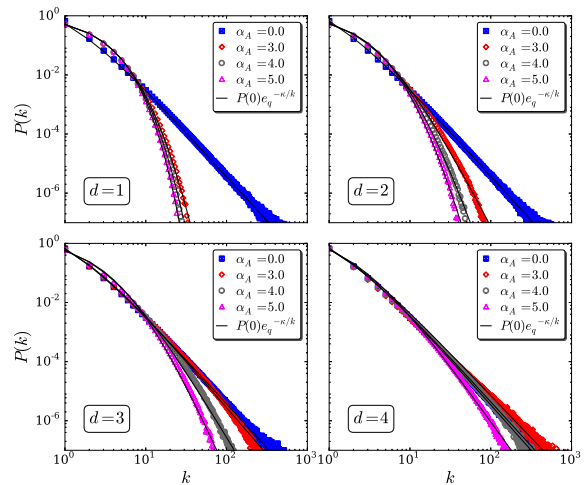}
\includegraphics[scale=.25]{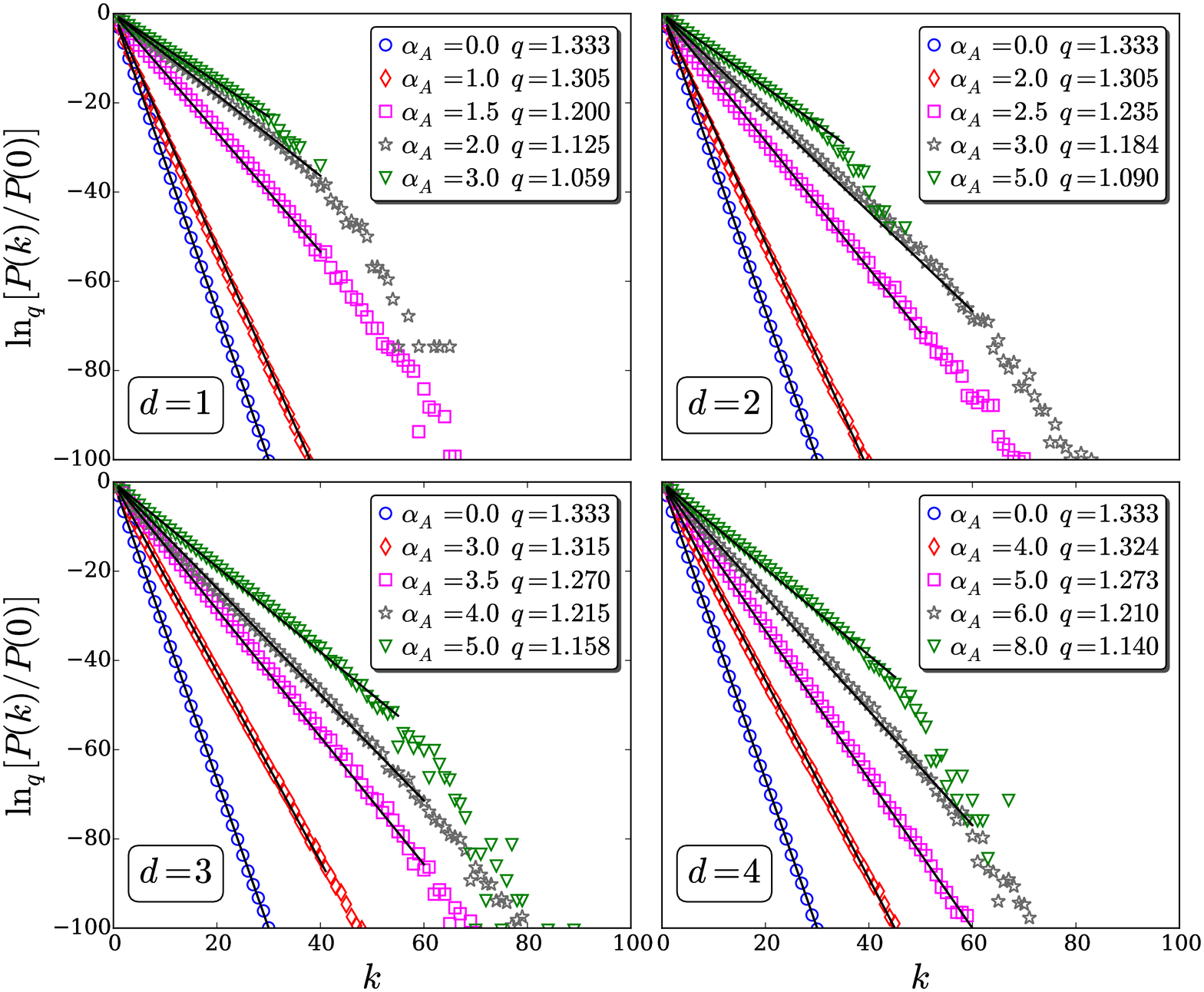}
\caption{Fittings of the $d=1,2,3,4$ connectivity distributions with the function $P(k) = P(0){e_q}^{-\kappa/k}$, where $e_q^z \equiv [1 + (1 - q) z]^{1/(1-q)}$. The data are those of Fig. \ref{compare}. {\it Top:} log-log representation. {\it Bottom:} $ln_q[P(k)/P(0)]$ versus $k$ representation. The fitting parameters are exhibited in Fig. \ref{q_kappa}. The numerical failure, at large enough values of $k$, with regard to straight lines are finite-size effects that gradually disappear when we approach the thermodynamic limit $N\to\infty$. }
\label{ajuste}
\end{figure}

\begin{figure}[!htb]
\centering
\includegraphics[scale=.34]{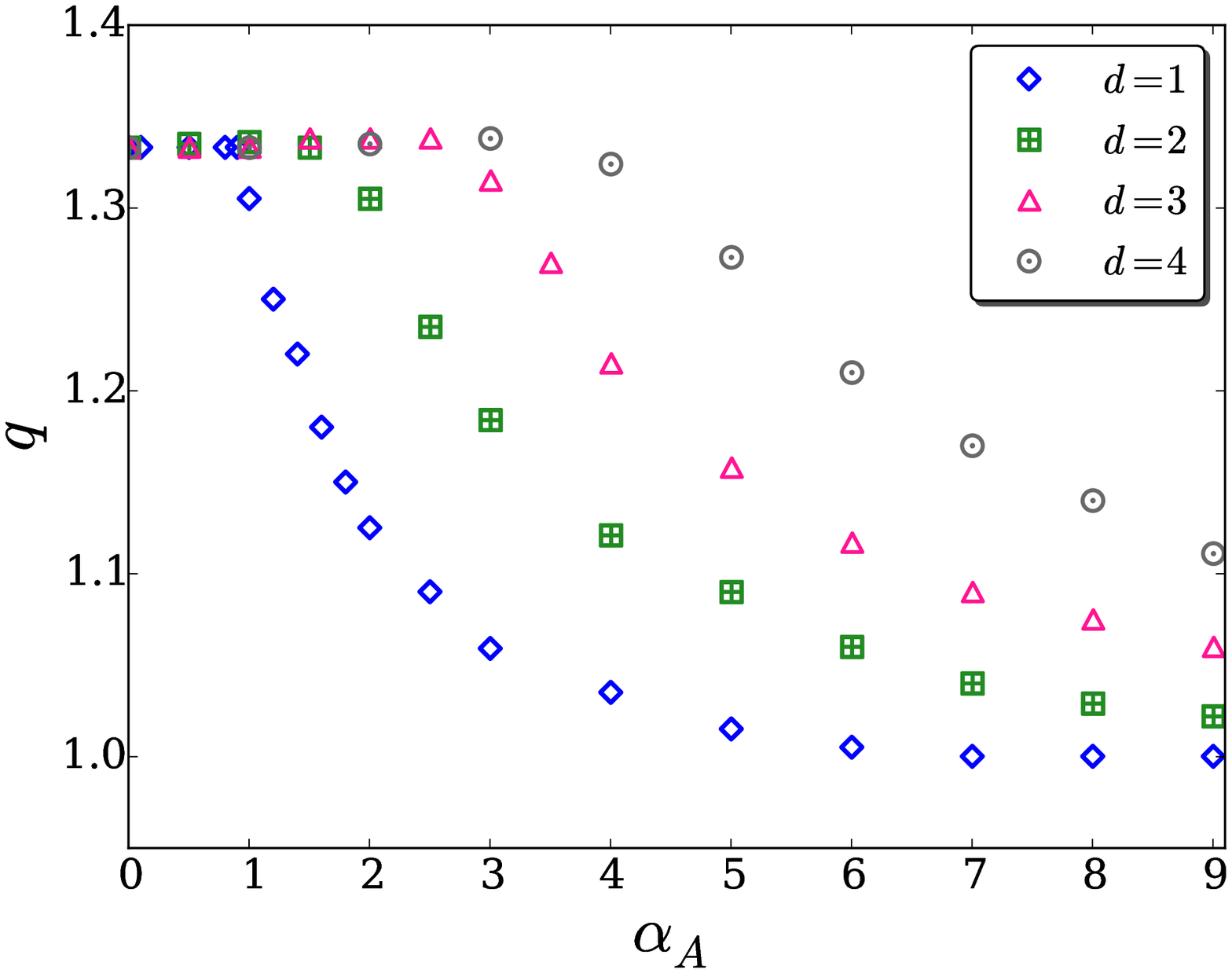}
\includegraphics[scale=.34]{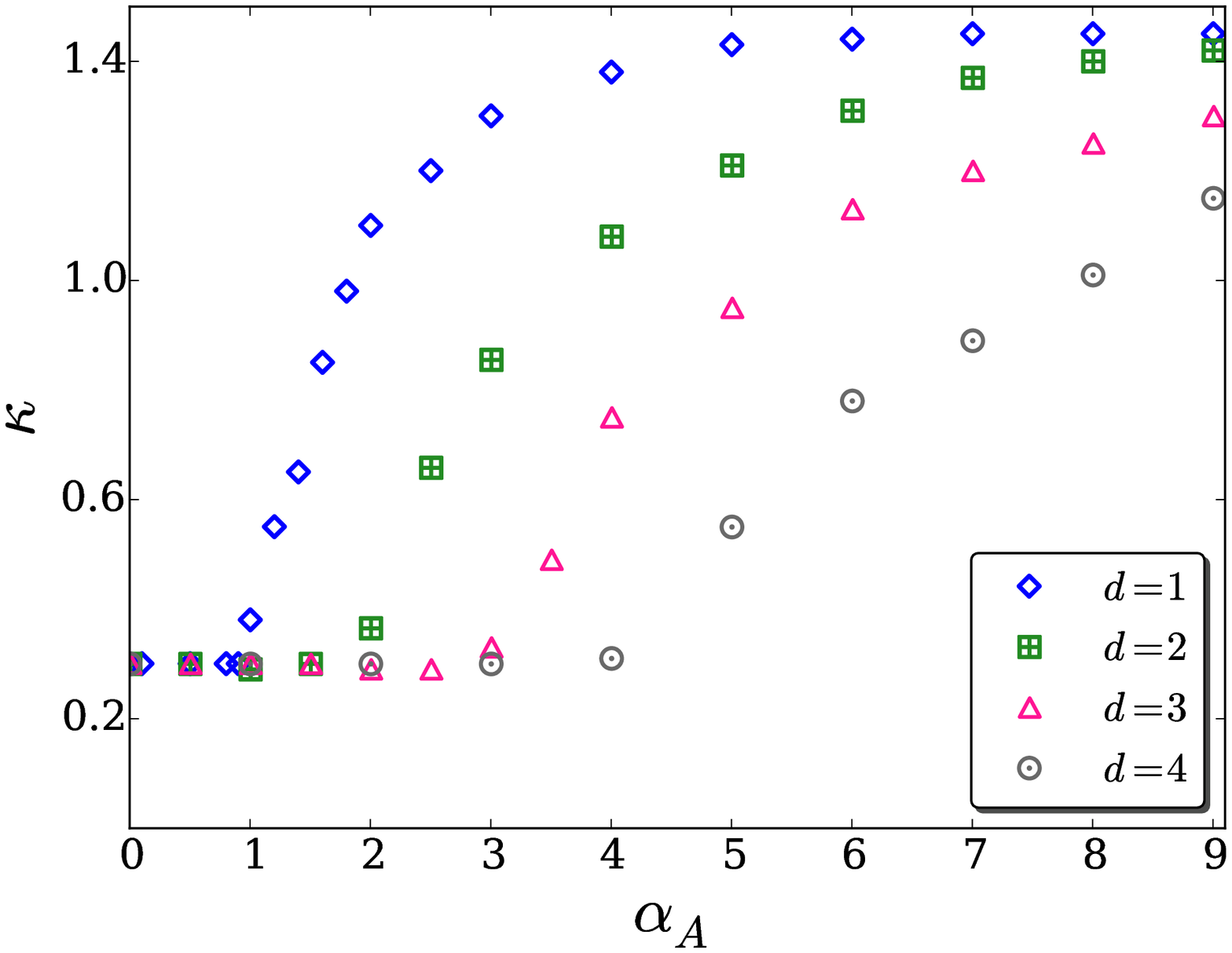}
\caption{$q$ and $\kappa$ for $d=1,2,3,4$. For $\alpha_A=0$ and $\forall d$, we recover the Barab\'asi-Albert universality class $q=4/3$ (hence $\gamma =3$) \cite{BarabasiAlbert1999}, which has no metrics.}
\label{q_kappa}
\end{figure}

Large-scale simulations have been performed for the $(d=1,2,3,4)$ models for fixed $(\alpha_G,\alpha_A)$, and we have verified in all cases that the degree distribution $P(k)$ is completely independent from $\alpha_G$: see Fig. \ref{alfa_g}. Using this fact, we have arbitrarily fixed $\alpha_G=2$, and have numerically studied the influence of $(d,\alpha_A)$ on $P(k)$: see Figs. \ref{compare} and \ref{ajuste}. In all cases, the $q$-exponential fittings $P(k)=P(0) e_q^{-k/\kappa}$ with $q>1$ and $\kappa>0$ have been remarkably good. The best fitting values for $(q,\kappa)$ are indicated in Fig. \ref{q_kappa}. From normalization of $P(k)$, $P(0)$ can be expressed as a straightforward function of $(q,\kappa)$. 

\begin{figure}[!htb]
\centering
\includegraphics[scale=.27]{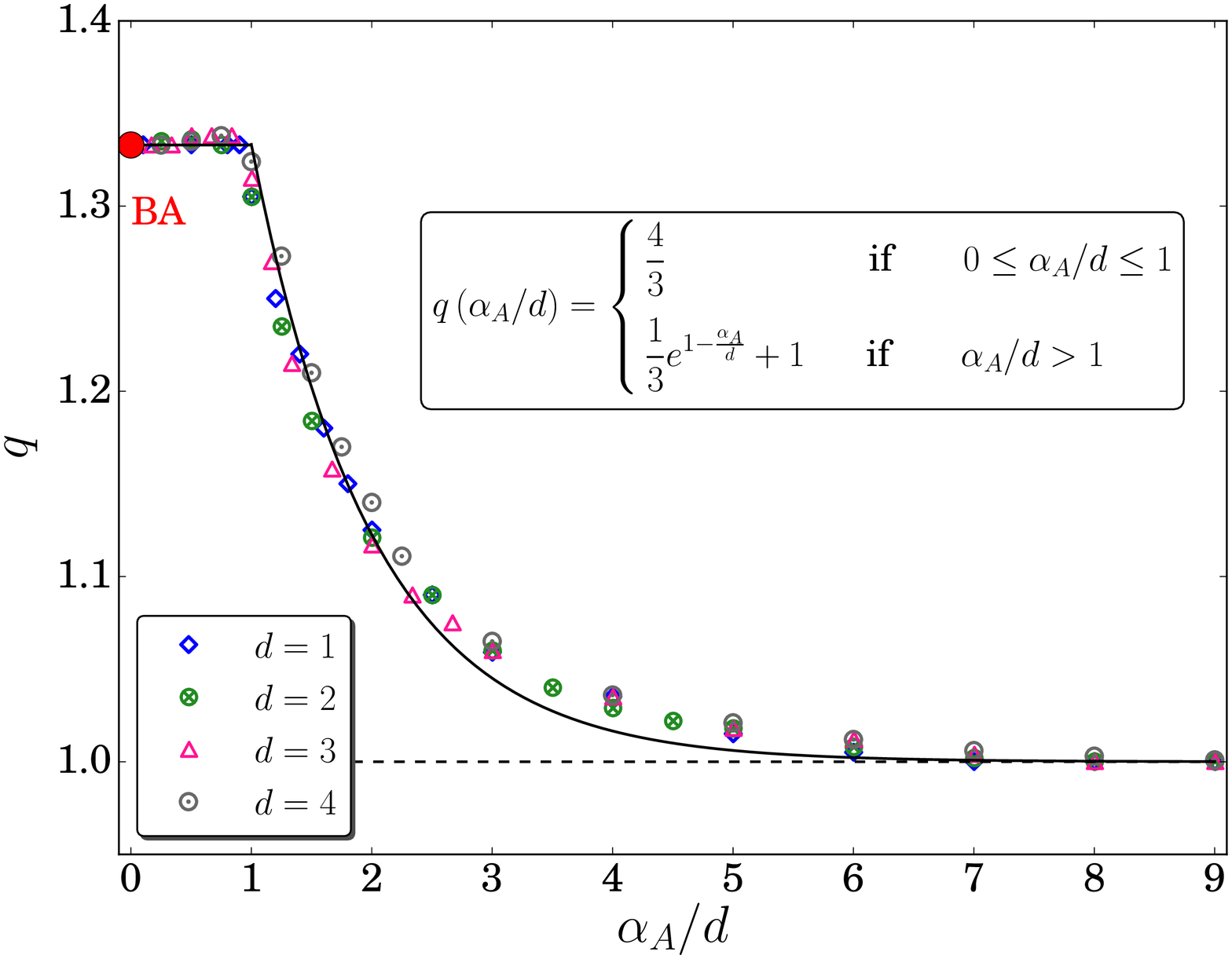}
\includegraphics[scale=.27]{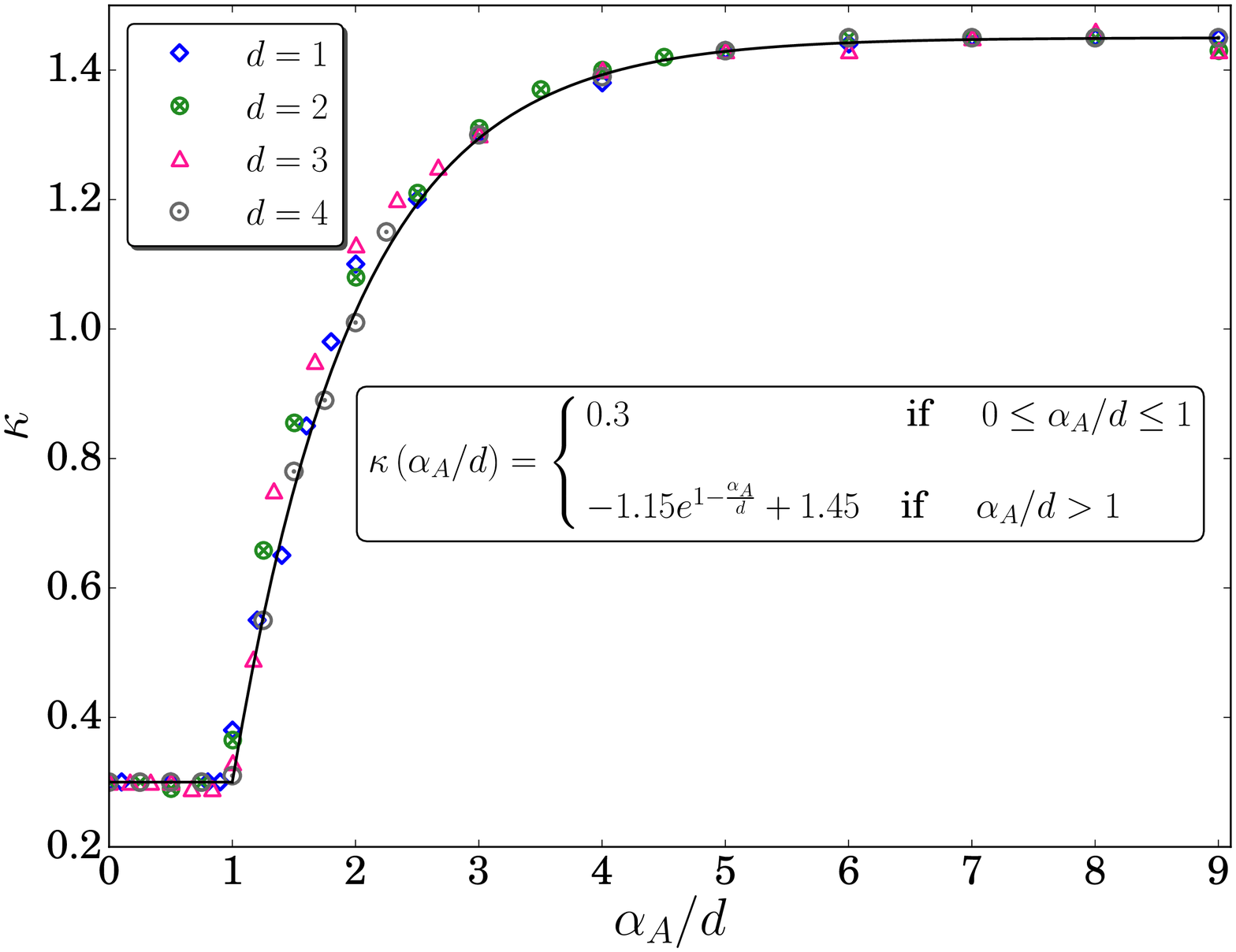}
\caption{$q$ and $\kappa$ versus $\alpha_A/d$ (same data as in Fig. \ref{q_kappa}). We see that $q = 4/3$ for $0 \le \alpha_A/d \leq 1$, and a nearly exponential behavior emerges for $\alpha_A/d > 1$ ($\forall d$); similarly for $\kappa$. These results exhibit the universality of both $q$ and $\kappa$. The red dot indicates the Barab\'asi-Albert (BA) universality class $q=4/3$ 
\cite{BarabasiAlbert1999}.}
\label{q_kapa_dim2}
\end{figure}

Our most remarkable results are presented in Fig. \ref{q_kapa_dim2}, namely the fact that both the index $q$ and the characteristic degree (or ``effective temperature") $\kappa$ do not depend from $(\alpha_A,d)$ in an independent manner but {\it only from the ratio $\alpha_A/d$}. This nontrivial fact puts the growing $d$-dimensional geographically located models that have been introduced here for scale-free networks, on similar footing as long-range-interacting many-body classical Hamiltonian systems such as the inertial XY planar rotators \cite{rotator} (possibly the generic inertial $n$-vector rotators as well \cite{heisenberg}) and Fermi-Pasta-Ulam \cite{FPU} oscillators, assuming that the strength of the two-body interaction decreases with distance as $1/(distance)^\alpha$. Moreover, as first pointed out generically by Gibbs himself \cite{gibbs}, we have the facts that the BG canonical partition function of these classical systems anomalously diverges with size for $0 \le \alpha/d \le 1$ (long-range interactions, e.g., gravitational and dipole-monopole interactions) and converges for $\alpha/d>1$ (short-range interactions, e.g., Lennard-Jones interaction), and the internal energy per particle is, in the thermodynamical limit, constant for short-range interactions whereas it diverges like $N^{1-\alpha/d}$ for long-range interactions, $N$ being the total number of particles.

If all these meaningful scalings are put together, we obtain a highly plausible scenario for the respective domains of validity of the Boltzmann-Gibbs (additive) entropy and associated statistical mechanics, and that of the nonadditive entropies $S_q$ (with $q \ne 1$) and associated statistical mechanics.

\begin{figure}[!htb]
\centering
\includegraphics[scale=.37]{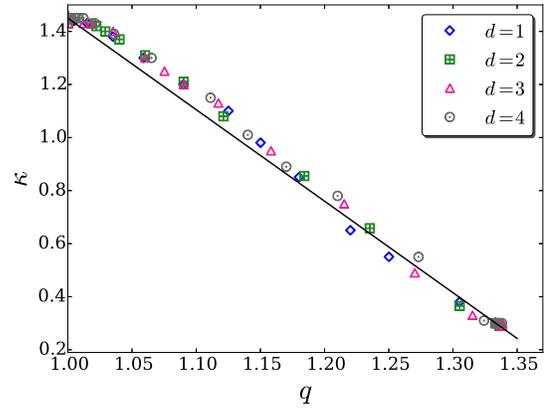}
\caption{All the values of $q$ and $\kappa$ for the present $d=1,2,3,4$  models follow closely the linear relation Eq. (\ref{relation}) (continuous straight line). The upmost value of $q$ is 4/3, yielding $\kappa \simeq 0.3$ ($\forall d$).}
\label{kapa_q}
\end{figure}

Finally, we notice in Fig. \ref{q_kapa_dim2} that both $q$ and $\kappa$ approach quickly their BG limits ($q=1$) for $\alpha_A/d \to\infty$. Moreover, the same exponential $e^{1-\alpha/d}$ appears in both heuristic expressions for $q$ and $\kappa$. Consequently, the following {\it linear}  relation can be straightforwardly established:
\begin{equation}
\kappa \simeq 4.90-3.45 \,q \,.
\label{relation}
\end{equation}
In fact, this simple relation is numerically quite well satisfied as can be seen in Fig. \ref{kapa_q}. Its existence reveals an interesting peculiarity of the nature of $q$-statistics. If in the celebrated BG factor $e^{-energy/kT}$, corresponding to $q=1$, we are free to consider an arbitrary value for $T$, how come in the present problem, $\kappa$ is not a free parameter but has instead a fixed value for each specific model that we are focusing on? This is precisely what occurs in the high-energy applications of $q$-statistics, e.g., in quark-gluon soup \cite{walton} where $q=1.114$ and $T=135.2 \,Mev$, as well as in all the LHC/CERN and RHIC/Brookhaven experiments \cite{cs9}. Another example which is reminiscent of this type of behavior is the sensitivity to the initial conditions at the edge of chaos (Feigenbaum point) of the logistic map; indeed, the inverse $q$-generalized Lyapunov exponent satisfies the linear relation  $1/\lambda_q = 1-q$ \cite{robledo}. 
The cause of this interesting and ubiquitous feature comes from the fact that $q$-statistics typically emerges at critical-like regimes and is deeply related to an hierarchical occupation of phase space (or Hilbert space or Fock space), which in turn points towards asymptotic power-laws (see also \cite{criticality}). In other words, $\kappa$ plays a role analogous to a critical temperature, which is of course not a free parameter but is instead fixed by the specific model.

\section*{Acknowledgments}
We have benefitted from  fruitful discussions with D. Bagchi, E.M.F. Curado, F.D. Nobre, P. Rapcan and G. Sicuro. We gratefully acknowledge partial financial support from CNPq and Faperj (Brazilian agencies) and from the John Templeton Foundation-USA.

\end{document}